# The Possible Source of the UHECR Observed in AGASA Experiment

Miroslaw Kozlowski[*], Janina Marciak-Kozlowska

Institute of Electron Technology, Al. Lotników 32/46, 02-668 Warsaw, Poland

Abstract

In this paper the results of AGASA experiment is discussed. It is argued that the UHECR radiation is composed of long lived particles with mass $m_x \approx 10^{13} m_\text{p}$, where $m_\text{p}$ is the proton mass.

---

[*] Corresponding author, e-mail: MiroslawKozlowski@aster.pl

## 1. Introduction

Cosmic Rays are observed in an energy range extending over more than eleven decades starting from sub-6 eV energies up to $\sim 10^{20}$ eV. The most prominent signature of extragaletic UHECR (ultrahigh energy cosmic rays) is so called GZK cutoff: the energy loses of protons increase sharply at $E_{GZK} \approx 5 \cdot 10^{19}$ eV, since pion production on cosmic microware background (CMB) photons $p + \gamma_{3K} \to \Delta^* \to N + \pi$ reduces their mean-free path by more than two orders of magnitude compared to lower energies [1].

The AGASA experiment [2] has detected a significant excess of events above $10^{20}$ eV. In this paper we will study of the AGASA UHECR spectrum in the spirit of the top-down model. The top-down model is a generic name for all proposals in which the observed UHECR primaries are produced as decay products of some superheavy particles $X$ with mass, $m_x > 10^{12}$ GeV. These $X$ particles can be either metastable or be emitted by topological defects at the present epoch.

The organization of the paper is as follows. After the short discussion of the New Physics models, paragraph 2, in paragraph 3 we develop the new Lorentz transformation with the change $c^2 \to -c^2$, c is the vacuum light speed. We will obtain the new formulae for energy, momentum, and kinetic energy of the new specic-particle $X$ with speed $v$, can be greater than $c$, $v_X > c$. It will be shown that the ratio $\left(v/c\right)^2$ is singular for $E_k \sim m_X c^2$. Starting from $\left(E_k / m_X c^2\right) > 10^{-2}$ the velocity of particle $X$ is different from the standard particles from $c^2$ Universe. Assuming that $\left(E_k / m_X c^2\right) \sim 10^{-2}$ is the threshold for observing particle $X$ and comparing it to post GZK part of the AGASA spectrum we obtain $m_X c^2 \sim 10^{24}\,\text{eV} \sim 10^{13}\,\text{GeV} \sim 10^{13} m_p$.

## 2. New Physics Models of UHECR

The UHECR puzzle has inspired a number of models that involve physics beyond the standard model of particle physics. New physics proposals can be top-down models or a hybrid of astrophysical Zevatrons ($10^{21}$ eV) with new particles. Top-down models involve the decay of very high mass relics that could have formed in the early universe or even before the Beginning. The most economical among hybrid proposals involves an extension of the standard model, namely, neutrino masses. If some flavor of neutrinos have mass ($\approx 0.1\,\text{eV}$), the relic neutrino background is a target for extremely high energy neutrinos to interact and generate other particles [1, 2]. In these proposals neutrino energies need to be $\geq 10^{21}$ eV which implies primary protons in the source with energies $\geq 10^{23}$ eV.

The idea behind the top-down models is that relicts of the very universe, topological defects (TDs) or superheavy relic (SHR) particles produced after or at the end of inflation can decay today and generate UHECRs.

Another interesting possibility is the proposal that UHECR are produced by the decay of unstable superheavy relics that live much longer than the age of universe [3, 4]. It was shown that these new particle species can emulate the observed UHECR fluxes provided 1) the new *X* particle is a very heavy neutrino-like particle, 2) tune $m_X$ to fit a description in which certain assumptions about decay are confined to a range – this becoming $m_X > 10^{13}\,\text{GeV} = 10^{22}\,\text{eV}$ and 3) assuming that the particle is stable on the timescale of the Universe.

In this paper we reconsider the top-down model assuming the following: 1) the AGASA energy spectrum breaks the GZK limit, 2) the spectrum can be described in the framework of the $-c^2$ universe in which $ic$ is invariant to the new Lorentz transformation.

## 3. The $-c^2$ Lorentz transformation

In monograph [5] we developed the hyperbolic heat transport equation

$$\tau \frac{\partial^2 T}{\partial t^2} + \frac{\partial T}{\partial t} = D \nabla^2 T. \tag{1}$$

In Eq. (1), $T$ is the temperature field $T(\vec{r},t)$, $\tau$ is the relaxation time, and $D$ denotes the diffusion coeficient. The transport coefficients $\tau$, and $D$, are defined viz.,

$$\tau = \frac{\hbar}{mv^2}, \qquad v = \alpha c, \qquad D = \tau v^2. \tag{2}$$

In formula (2) $m$ is the mass of heat carrier, $v$ is the speed of thermal disturbance propagation, and $\alpha$ is coupling constant of the interaction (electromagnetic, coloured or gravitation). Let us perform the transformation

$$c \to ic. \tag{3}$$

We observe that (for electomagnetic interaction)

$$\alpha_T = \frac{e^2}{\hbar ic} = -i\alpha, \tag{4}$$

$$\begin{aligned} v_T &= \alpha_T (ic) = v, \\ \tau_T &= \frac{\hbar}{mv_T^2} = \tau, \\ D_T &= D \end{aligned} \tag{5}$$

where subscript T denotes the transformed, for $c^2 \to -c^2$, values of the transport coefficients. From formulae (3)–(5) we conclude that Eq. (1) is invariant against the transformation (3).

Let us consider the Lorentz formulae for inertial frames $(x, t)$ and $(x',t')$

$$x' = \gamma(x - vt), \qquad t' = \gamma\left(t - \frac{v}{c^2}x\right), \tag{6}$$

$$\gamma = \frac{1}{\sqrt{1 - \frac{v^2}{c^2}}}.$$

In the $-c^2$ universe we obtain instead of (6)

$$\gamma' = \frac{1}{\sqrt{1+\frac{v^2}{c^2}}}, \tag{7}$$

$$x' = \gamma'(x-vt), \qquad t' = \gamma'\left(t + \frac{v}{c^2}x\right).$$

For space-time interval we obtain

$$(x')^2 + (ct')^2 = x^2 + (ct)^2.$$

For composition of velocities we obtain new formula

$$v' = \frac{v-V}{1+\frac{vV}{c^2}}. \tag{8}$$

For special velocity $v = ic$ we obtain from formula (8)

$$v' = \frac{ic-V}{1+\frac{icV}{c^2}} = ic. \tag{9}$$

From formula (9) we conclude that $ic$ is the same in all inertial coordinate frames.

In $-ic^2$ universe the total energy of the particle is defined as follows

$$E = -\frac{mc^2}{\sqrt{1+\frac{v^2}{c^2}}}, \tag{10}$$

and kinetic energy

$$E_k = -mc^2(\gamma'-1), \qquad \gamma' = \frac{1}{\sqrt{1+\frac{v^2}{c^2}}}, \tag{11}$$

and momentum of the particle

$$p = mv\gamma'. \tag{12}$$

From formulae (10) and (12) one obtains

$$\frac{E}{p} = -\frac{c^2}{v}; \quad v = -\frac{c^2 p}{E}. \tag{13}$$

For particle with velocity $ic$ we obtain

$$E = ipc, \quad \text{i.e.} \quad m = 0. \tag{14}$$

For particle with $m \neq 0$ we obtain

$$E^2 = -p^2 c^2 + m^2 c^4. \tag{15}$$

From formula (11) we calculate the ratio $v^2/c^2$, i.e.

$$\frac{v^2}{c^2} = \frac{1}{\left(1 - \frac{E_k}{mc^2}\right)} - 1. \tag{16}$$

As can be realized from formula (16), $v^2/c^2 \to \infty$, for $E_k \to mc^2$. In Fig. 1 we present $v^2/c^2$ as the function of the ratio $E_k/mc^2$. From Figs. 1 and 2 we conclude that particles in the $-c^2$ universe, and particles in $c^2$ behave in the same manner only for $E_k/mc^2 \ll 0.02$. As was stated early for $E_k == mc^2$, $v^2 \to \infty$ On the other hand for $E_k > mc^2$ particles behaves as the tachyons, i.e. for increasing $E_k$ speed of particle is decreasing.

## 4. The AGASA results

From the results of the AGASA experiment, i.e. energy spectra of UHECR we conclude. that for $E_k \approx 10^{20}$ the shape of the spectra change rather abruptly. From the comparison of Fig. 1 and Fig. 3 we argue that the change of the spectrum is due to the fact that we observe the singularity due to the $E_k \to mc^2$ where $m$ is the mass of the decaying particle. We argue that

$$\frac{E_k}{mc^2} \approx 0.02 \tag{17}$$

can be used to the calculation of the mass of the decaying particle

$$mc^2 \approx 10^2 E_k. \tag{18}$$

Considering the results of AGASA experiment we obtain

$$M = mc^2 \approx 10^{22}\,\text{eV} \approx 10^{13}\,\text{GeV} \approx 10^{13}\,m_N c^2.$$

Now we calculate the life-time the particle with mass $M \approx 10^{13} m_N$. From formula (7) we obtain

$$\tau_M = \tau\gamma'; \qquad \gamma' < 1 \tag{19}$$

In formula (19), $\tau$ is the life-time of the particle in its inertial reference frame, and $\tau_M$ is the life-time of the particle in the Earth reference frame. For the Earth observer $\tau_M \approx 10^{17}$ s and from formula (19) we obtain

$$\tau = 10^{17}\sqrt{1 + \frac{v^2}{c^2}}. \tag{20}$$

For $E_k \approx mc^2$, $v/c^2 \to \infty$, and $\tau > 10^{17}$ s i.e. the life-time of particle $M$ is greater than the life-time of our Universe.

,

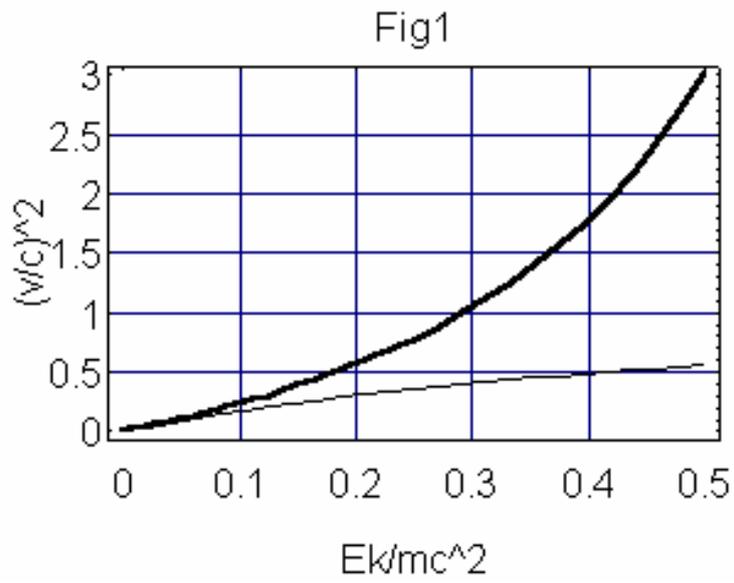

Fig. 1. The $\left(v/c\right)^2$ as the function of the $E_k/mc^2$.

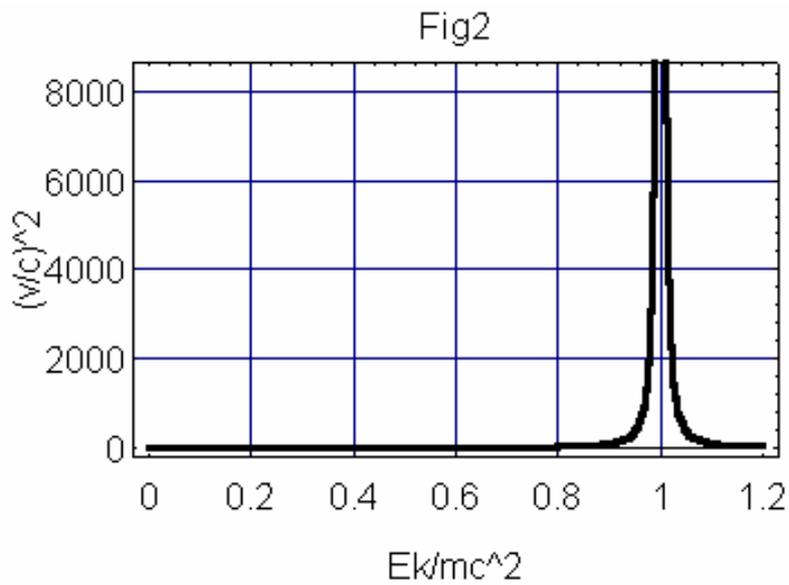

Fig. 2. The same as in Fig. 1 but for large values of $E_k/mc^2$.

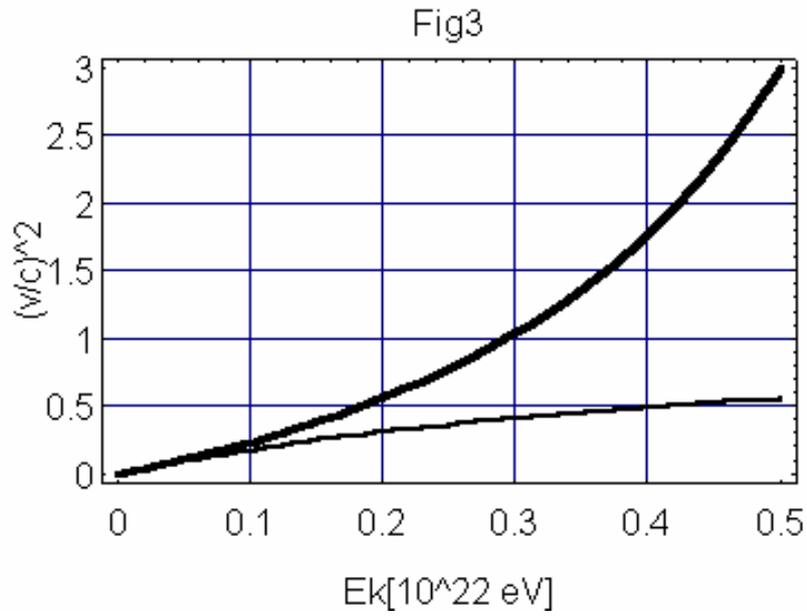

Fig. 3. The $\left(v/c\right)^2$ as the function of $E_k$ observed in AGASA experiment.

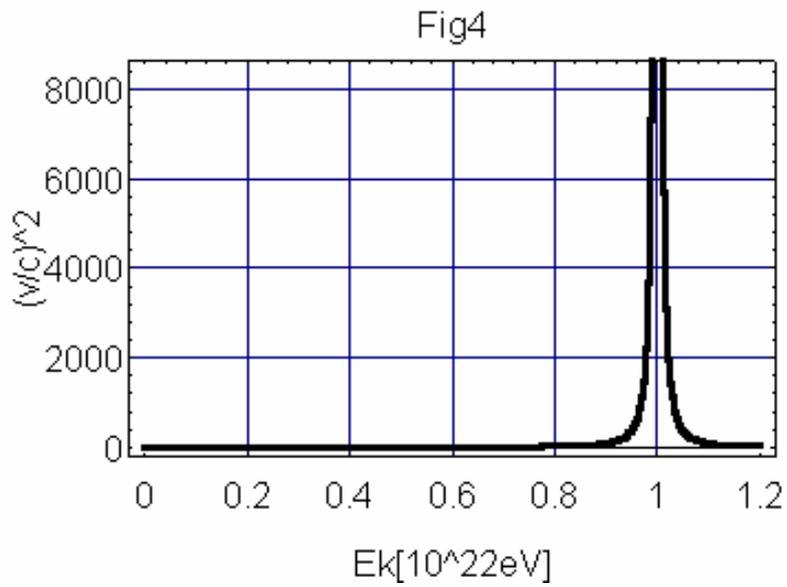

Fig. 4. The same as in Fig. 4 but for large values of $E_k$.